\documentclass[nofootinbib,twocolumn,prd,preprintnumbers,superscriptaddress,aps]{revtex4}
\usepackage[utf8]{inputenc}
\pdfoutput=1 

\usepackage{comment} 
\usepackage{graphicx} 
\usepackage{epsfig}
\usepackage{bm}
\usepackage{amssymb}
\usepackage{float}
\usepackage{amsmath}
\usepackage{dcolumn}
\usepackage{cancel}
\usepackage[colorlinks]{hyperref}
\usepackage[usenames, dvipsnames]{color}

\hypersetup{
     breaklinks=true, 
    pdfstartview={FitH},  
    colorlinks=true, 
    linkcolor=blue,  
    citecolor=red,  
    filecolor=magenta,  
    urlcolor=blue, 
    anchorcolor=green,  
    linktocpage=true
}

\providecommand{\U}[1]{\protect\rule{.1in}{.1in}}

\newcommand{\be}{\begin{equation}}
\newcommand{\ee}{\end{equation}}

\newcommand{\mincir}{\raise
-3.truept\hbox{\rlap{\hbox{$\sim$}}\raise4.truept\hbox{$<$}\ }}
\newcommand{\magcir}{\raise
-3.truept\hbox{\rlap{\hbox{$\sim$}}\raise4.truept\hbox{$>$}\ }}

\ifx\pdfoutput\relax\let\pdfoutput=\undefined\fi
\newcount\msipdfoutput
\ifx\pdfoutput\undefined\else
\ifcase\pdfoutput\else
\ifx\paperwidth\undefined\else
\ifdim\paperheight=0pt\relax\else\pdfpageheight\paperheight\fi
\ifdim\paperwidth=0pt\relax\else\pdfpagewidth\paperwidth\fi
\fi\fi\fi

\hypersetup{colorlinks=true,
	breaklinks=true,
	pdfstartview=Fit,
	linkcolor=blue,
	citecolor=blue,
	urlcolor=blue}

\begin{document}
\title{Barrow and Tsallis entropies after the DESI DR2 BAO data}

\author{Giuseppe Gaetano Luciano}
\email{giuseppegaetano.luciano@udl.cat}
\affiliation{Departamento de Qu\'{\i}mica, F\'{\i}sica y Ciencias Ambientales y del Suelo, Escuela Polit\'ecnica Superior -- Lleida, Universidad de Lleida, Av. Jaume II, 69, 25001 Lleida, Spain}

\author{Andronikos Paliathanasis}
\email{anpaliat@phys.uoa.gr}
\affiliation{Department of Mathematics, Faculty of Applied
Sciences, Durban University of Technology, Durban 4000, South Africa}
\affiliation{School for Data Science and Computational Thinking, Stellenbosch 
University,44 Banghoek Rd, Stellenbosch 7600, South Africa}
\affiliation{Centre for Space Research, North-West University, Potchefstroom 
2520, South Africa}
\affiliation{Departamento de Matem\`{a}ticas, Universidad Cat\`{o}lica del 
Norte, Avda.
Angamos 0610, Casilla 1280 Antofagasta, Chile}

\author{Emmanuel N. Saridakis}
\email{msaridak@noa.gr}
\affiliation{National Observatory of Athens, Lofos Nymfon 11852, Greece}
\affiliation{CAS Key Laboratory for Research in Galaxies and Cosmology, School 
of Astronomy and Space Science, University of Science and Technology of China, 
Hefei 230026, China}
\affiliation{Departamento de Matem\'{a}ticas, Universidad Cat\'{o}lica del 
Norte, Avda. Angamos 0610, Casilla 1280, Antofagasta, Chile}

\begin{abstract} 
Modified cosmology based on Barrow entropy arises from the 
gravity–thermodynamics conjecture,  in which the standard Bekenstein–Hawking 
entropy is replaced by the Barrow entropy of quantum-gravitational origin, 
characterized by the Barrow parameter $\Delta$. Interestingly, this framework 
exhibits similarities with cosmology based on Tsallis $\delta$-entropy, which, 
although rooted in a non-extensive generalization of Boltzmann–Gibbs statistics, 
features the same power-law deformation of the holographic scaling present in 
the Barrow case.
We use observational data from Supernova Type Ia (SNIa), Cosmic 
Chronometers (CC), and Baryonic acoustic oscillations (BAO), including the 
recently released DESI  DR2 data, in order to extract constraints on such 
scenarios. As we show,   the best-fit  value for the Barrow exponent  $\Delta$ 
is found to be negative, while the zero value, which corresponds to 
$\Lambda$CDM paradigm,   is allowed only   in the range of $2\sigma $ for three 
out of four datasets. Additionally, for the case of the   
 SN$_{0}$+OHD+BAO dataset, for the current Hubble function we 
obtain a value of $H_0= 72.2_{-0.9}^{+0.9}$, which may 
potentially alleviate the $H_0$ tension. Moreover,  we 
compare our 
results with  the Gaussian Process reconstruction of the dark-energy equation 
of 
state for the DESI + Cosmic Microwave Background  (CMB) + Union3 datasets.  
Finally, by applying information criteria such as the  
Akaike Information Criterion and the   Bayes evidence, we compare the fitting 
efficiency of the scenario at hand with $\Lambda$CDM  cosmology, showing that 
the latter is slightly favoured.

\end{abstract}

\maketitle

\section{Introduction}
One of the central open problems  in modern theoretical physics is the 
unification of quantum mechanics with general relativity. Despite substantial 
progress over recent decades, a complete and consistent theory of quantum 
gravity remains elusive. Various frameworks have been proposed, yet each faces 
its own technical limitations or conceptual 
challenges~\cite{Rovelli:2000aw,Addazi:2021xuf}. In this context, black holes 
serve as a crucial theoretical laboratory, offering profound opportunities to 
explore the interplay between gravity, quantum theory and thermodynamics. Their 
unique properties make them promising candidates for probing the fundamental 
principles underlying a unified description of nature.

In 2020, Barrow proposed that the Bekenstein-Hawking entropy may be subject to  
a quantum-level deformation, deviating from the standard holographic 
scaling~\cite{Barrow:2020tzx}. This modification arises from the hypothesis 
that 
quantum fluctuations induce a fractal-like geometry on the surface of a black 
hole, or more generally, on the area of any holographic 
horizon~\cite{Barrow:2020tzx,JalalzadehI,JalalzadehII}. As a result, the 
entropy 
receives a correction characterized by a deviation parameter $\Delta$, which 
accounts for the degree of quantum-induced fractality. Specifically, the Barrow 
entropy scales as
\begin{equation}
S_B  =   \left(\frac{A}{A_0}\right)^{1+\Delta/2}\, ,
\label{Barrow}
\end{equation}
where $A = L^2$ denotes the standard horizon area and $A_0=4L_p^2$ represents 
the Planck area.  The exponent $\Delta$ quantifies the degree of 
quantum-gravitational deformation and lies within the interval $0 \leq \Delta 
\leq 1$. The extreme case $\Delta = 1$ corresponds to a maximally intricate, 
fractal-like horizon geometry, while $\Delta = 0$ reflects a perfectly smooth 
configuration, thereby recovering the classical Bekenstein-Hawking entropy. 

It is worth noting that the entropy expression in \eqref{Barrow} 
exhibits a formal resemblance to the Tsallis  $\delta$-entropy, given by 
$S_\delta \propto A^\delta$, provided one identifies $\Delta \rightarrow 
2(\delta - 1)$~\cite{Tsallis:1987eu,Tsallis:2009,Tsallis:2013,Jizba:2022bfz}. 
However, despite this mathematical correspondence, the two entropy expressions have fundamentally distinct physical motivations and interpretations. 
Tsallis entropy arises in the context of nonextensive statistical mechanics, which generalizes the standard Boltzmann-Gibbs formalism to systems with memory effects,  multifractal structures or long-range interactions, such as gravitational systems. In this sense, nonextensive effects may naturally emerge as relevant even at cosmological scales, thereby providing a solid theoretical motivation for applying Tsallis entropy to the thermodynamical description of the Universe as a whole.
In all such systems, the assumption of additivity, which ensures the 
extensivity of Boltzmann-Gibbs entropy, no longer holds. Tsallis proposed a new 
entropy measure to account for such a possibility, thereby enabling a 
more accurate thermodynamical and statistical description of complex systems. 
Notably, the Legendre structure of thermodynamics is preserved in the Tsallis 
formalism, provided one defines  
expectation values  and internal energy consistently~\cite{Tsallis:2009}. In the 
following, we will explicitly develop our analysis within the Barrow entropy 
model. However, in light of the above discussion, it is clear that the results 
and considerations obtained can be mathematically applied to the Tsallis 
framework as well.

While Barrow originally motivated the parameter $\Delta$ through a  simple 
fractal model - specifically, a “sphereflake” construction - which naturally 
restricts $\Delta$ to values within the interval $[0,1]$, more general 
considerations indicate that this range could be extended to include negative 
values. In particular, surface geometries characterized by voids or internal 
porosity, such as sponge-like or porous structures, can exhibit effective 
fractal dimensions significantly lower than the embedding Euclidean dimension. 
For instance, the Sierpi{\'n}ski carpet has a Hausdorff dimension of 
approximately $1.89$, implying a corresponding value of $\Delta \approx -0.11$. 
Empirical studies have reported even lower effective dimensions in real-world 
porous materials, in some cases approaching values close to $1$ (i.e., $\Delta 
\approx -1)$~\cite{Tang,Xu}. 
Moreover, support for $\Delta < 0$ also emerges from theoretical arguments  
within quantum field theory, where renormalization group flow can give rise to 
negative anomalous dimensions in certain systems~\cite{Kogut}. These insights 
motivate a broader consideration of the parameter space, allowing for $\Delta 
\in (-1,1]$ in a more general 
framework~\cite{Jizba:2024klq,Anagnostopoulos:2020ctz}.

In the regime of small $\Delta$, which is anticipated  to be the physically 
relevant case, the Barrow entropy $S_B$ can be approximated as $
S_B =  
\frac{A}{A_0}\left[1+\frac{\Delta}{2}\,\mathrm{Log}\left(\frac{A}{A_0}
\right)\right] + \mathcal{O}(\Delta^2)$. Interestingly, logarithmic corrections 
to black hole entropy are a common prediction across a wide range of quantum 
gravity theories, such as string theory~\cite{Banerjee:2011jp}, loop quantum 
gravity~\cite{Kaul:2000kf}, the AdS/CFT correspondence~\cite{Carlip:2000nv} and 
models incorporating generalized uncertainty principles~\cite{Adler:2001vs}. 
Similarly, corrections to the Bekenstein entropy computed via the Cardy 
formula~\cite{Carlip:2000nv} exhibit a logarithmic behavior (with a negative 
$\Delta$) in several frameworks, including models based on asymptotic 
symmetries, horizon symmetries and certain formulations of string theory.
Therefore, although our interpretation associates the entropy deformation with 
quantum-induced fluctuations of the horizon geometry, the resulting structure is 
likely to be robust and applicable within a broader class of quantum gravity 
scenarios.

The modified cosmology  through Barrow entropy  has been recently extended 
to cosmological settings, providing a modified thermodynamic foundation for the 
evolution of the 
Universe~\cite{Barboza:2014yfe,Nunes:2015xsa,Barrow:2020tzx,Saridakis:2020zol,Nojiri:2021jxf,Barrow:2020kug, 
Saridakis:2020lrg,Adhikary:2021xym,Sheykhi:2021fwh,DiGennaro:2022ykp, 
Dabrowski:2020atl,Mamon:2020spa,Jusufi:2021fek,Luciano:2022pzg,Luciano:2022hhy, 
Luciano:2023wtx,Jizba:2024klq,Luciano:2025fox}. This line of research is 
motivated by the gravity–thermodynamics conjecture, which posits a deep 
connection between gravitational dynamics and thermodynamic laws, particularly 
in the context of horizon 
thermodynamics~\cite{Jacobson:1995ab,Padmanabhan:2009vy}. Within this 
framework, 
modifications to the entropy–area relation lead to generalized Friedmann 
equations, offering a new perspective on cosmic 
evolution~\cite{Luciano:2022ffn} 
and potentially addressing observational tensions in the $\Lambda$CDM 
paradigm~\cite{Basilakos:2023kvk,Luciano:2023roh,Yarahmadi:2024oqv}. 

In a  cosmological context, it is also natural to consider the possibility that 
the deformation parameter $\Delta$ evolves dynamically with the energy scale, 
rather than being treated as a fixed constant as in Barrow's original 
formulation. This idea aligns with the behavior of the effective gravitational 
coupling in various approaches to quantum gravity, where it typically weakens 
at 
large distances or low energies~\cite{tHooft:1993dmi}. Consequently, one 
expects 
$\Delta$ to decrease in the infrared regime, asymptotically approaching zero as 
quantum gravitational effects become negligible, thereby recovering the 
classical Bekenstein--Hawking entropy. Investigations along this direction have 
been recently pursued in~\cite{DiGennaro:2022ykp,Basilakos:2023seo}, further 
enriching the theoretical landscape of entropic cosmology.

On the other hand, the recent release of DESI and DESI DR2 has already demonstrated its utility as a powerful tool for constraining a broad class of cosmological models~\cite{DESI:2025zgx}. Owing to the unprecedented precision of Baryon Acoustic Oscillation (BAO) measurements, these data have been employed to rigorously test a variety of extensions to the standard cosmological model, including dynamical dark energy scenarios~
\cite{Ormondroyd:2025iaf,You:2025uon,Gu:2025xie,Santos:2025wiv,Li:2025cxn,Alfano:2025gie,Carloni:2024zpl},
early dark energy \cite{Chaussidon:2025npr}, scalar field theories with both 
minimal and non-minimal 
couplings~\cite{Anchordoqui:2025fgz,Ye:2025ulq,Wolf:2025jed}, models inspired 
by 
the Generalized Uncertainty Principle~\cite{Paliathanasis:2025dcr}, interacting 
dark sector scenarios~\cite{Shah:2025ayl,Silva:2025hxw,Pan:2025qwy}, astrophysical models~\cite{Alfano:2024jqn}, cosmographic analysis~\cite{Luongo:2024fww} and several 
modified theories of gravity~\cite{Yang:2025mws,Li:2025cxn,Paliathanasis:2025hjw,Tyagi:2025zov}. 

In the present work, we aim to employ observational data from the recent DESI 
DR2 release in order to constrain the Barrow cosmological framework. In 
particular, we focus on deriving observational bounds on the Barrow exponent, 
which quantifies the degree of quantum-gravitational deformation and thereby 
encapsulates deviations from the standard cosmological model. The remainder of 
this work is organized as follows. In the next section, we review the modified 
Friedmann equations that arise from applying the gravity–thermodynamics 
correspondence within the Barrow entropy framework. In Section~\ref{Obs}, we 
present the cosmological constraints on the entropic model, using the BAO 
measurements from DESI DR2 combined with the Pantheon+ Supernova dataset and 
Cosmic Chronometers. Finally, in Section~\ref{Conc}, we summarize our findings 
and present our conclusions. An appendix contains mathematical details regarding the derivation of the Friedmann equations within the extended entropy model.

\section{Modified Cosmology Driven by Barrow or Tsallis Entropy}
\label{ModCosm}
In this section, we provide an overview of the derivation of the modified 
Friedmann equations within the framework of the gravity-thermodynamics 
conjecture, incorporating the Barrow entropy in place of the standard area-law 
entropy. Following the methodology outlined in 
Refs.~\cite{Saridakis:2020lrg,Barrow:2020kug,Luciano:2022pzg}, we interpret the 
contributions arising from such an extended entropy as corrections to the total 
energy density in the Friedmann equations.  
To set the stage and establish notation, we begin by briefly reviewing the 
standard derivation of the cosmological equations in the case $\Delta = 0$, 
considering a spatially flat Friedmann–Robertson–Walker (FRW) Universe. 

\subsection{Friedmann equations from thermodynamic laws}

The formal analogy between the laws of black-hole mechanics and classical 
thermodynamics has long motivated the development of black-hole thermodynamics, 
initially without reference to statistical mechanics~\cite{Bardeen}. This 
framework was significantly extended by Gibbons and 
Hawking~\cite{Gibbons:1977mu} and later by 't Hooft~\cite{tHooft:1993dmi} and 
Susskind~\cite{Susskind:1994vu}, who demonstrated that thermodynamic properties 
- such as entropy and temperature - are associated not only with black holes, 
but also with general event horizons. This insight was further reinforced by the 
AdS/CFT correspondence, which revealed that the entropy of the boundary 
conformal field theory is directly related to the horizon area of the bulk AdS 
black hole~\cite{Ryu:2006ef,Aharony:2008ug}.

The above developments provide compelling motivation for applying thermodynamic 
principles to cosmological frameworks. In what follows, we adopt this approach 
to derive the Friedmann equations by implementing the first law of 
thermodynamics at the apparent horizon of a spatially flat, homogeneous and isotropic FRW universe, characterized by the line element
\begin{equation}
ds^2 =  h_{\mu\nu} \, dx^\mu dx^\nu \ + \ \tilde{r}^2\left(d\theta^2 \ + \ 
\sin^2\theta\, d\phi^2\right)\, ,
\label{FRW}
\end{equation}
where the areal radius is defined as $\tilde{r} = a(t)\hspace{0.2mm}r$, with 
coordinates $x^0 = t$ and $x^1 = r$. The metric $h_{\mu\nu}$ corresponds to the 
two-dimensional submanifold spanned by the coordinates $(t, r)$ and takes the 
form $h_{\mu\nu} = \mathrm{diag}(-1, a^2)$, where $a(t)$ is the time-dependent 
scale factor. We further assume that the matter content of the Universe can be 
effectively described by a perfect fluid, which is consistent with the 
symmetries of the FRW spacetime and widely adopted in standard cosmological 
models.

In this setup, the dynamical apparent horizon plays a crucial role in defining 
thermodynamic quantities. For a spatially flat FRW universe, it is given by 
$\tilde{r}_A = 1/H$~\cite{Frolov:2002va,Cai:2005ra,Cai:2009qf},
where $H = \dot{a}/a$ is the Hubble parameter (the overdot denotes time 
derivative).  In turn, the temperature at the apparent horizon is typically 
taken to be the Hawking-like temperature
$T_h = 1/(2\pi \tilde{r}_A)$,
in analogy with black-hole thermodynamics \cite{Cai:2009qf,Padmanabhan:2009vy}. 
In this context, we adopt the hypothesis of a quasi-static expansion of the 
Universe~\cite{Luciano:2023zrx}, which ensures that the horizon temperature 
remains well-defined throughout cosmic evolution. Furthermore, we assume that 
the cosmic fluid is in thermal equilibrium with the apparent horizon, as a 
consequence of long-term 
interactions~\cite{Padmanabhan:2009vy,Frolov:2002va,Cai:2005ra,Izquierdo:2005ku,
Akbar:2006kj}. This equilibrium condition justifies the use of equilibrium 
thermodynamics and allows us to bypass the mathematical complications associated 
with non-equilibrium frameworks.

The analogy between black-hole thermodynamics and cosmology can be further 
extended by introducing the notion of apparent horizon entropy. Within the 
framework of standard General Relativity, this entropy is determined by the 
Bekenstein–Hawking formula, expressed as $S_H = A / A_0$, where $A = 4\pi 
\tilde{r}_A^2$ denotes the surface area of the apparent horizon and $A_0$ 
represents the Planck area defined below Eq.~\eqref{Barrow}.
 
At this point, we invoke the first law of black hole thermodynamics, which can 
be cast in the form
\begin{equation}
dU = T_h\, dS - \mathcal{W}\, dV\,,
\label{14c}
\end{equation}
where $\mathcal{W}$ represents the work density, playing a role analogous to 
pressure in fluid dynamics, while $V = 4\pi\tilde{r}_A^3/3$ denotes the total 
volume enclosed by the apparent horizon, effectively corresponding to the 
observable portion of the Universe at a given time.

Under the perfect fluid assumption introduced earlier, the energy-momentum 
tensor describing the matter content of the Universe takes the standard form
\begin{equation}
T_{\mu\nu} = (\rho + p)u_{\mu}u_{\nu} + p\,g_{\mu\nu}\,,
\end{equation}
where $\rho$ is the energy density, $p$ is the isotropic pressure and $u^\mu$ 
denotes the four-velocity of the fluid. In this framework, the conservation of 
energy–momentum leads to the continuity equation, namely
\begin{equation}
\label{cont}
\nabla_\mu T^{\mu\nu} = 0 \,\,\, \Longrightarrow\,\,\, \dot{\rho} + 3H(\rho + p) 
= 0\,,
\end{equation}
which governs the evolution of the energy density as the Universe expands. The 
corresponding work density, which arises due to variations in the apparent 
horizon radius, is given by $\mathcal{W} = -\frac{1}{2} \, \text{Tr}(T^{\mu\nu}) 
= \frac{1}{2}(\rho - p)$, where the trace is taken with respect to the induced 
metric on the two-dimensional $(t, r)$ submanifold, i.e., $\text{Tr}(T^{\mu\nu}) 
= T^{\alpha\beta} h_{\alpha\beta}$.

As the Universe evolves over an infinitesimal time interval $dt$, the change in 
internal energy $dU$ associated with the variation of the apparent horizon 
volume corresponds to a reduction in the total energy $E$ contained within that 
volume. This leads to the relation $dU = -dE$, which implies 
\begin{equation}
    dE \ = \ - T_hdS \ + \ \mathcal{W}dV\, .
\end{equation}
By making use of the relation $E = \rho V$ and combining it with the continuity 
equation~\eqref{cont}, one obtains 
\begin{equation}
T_h\, dS = A\left(\rho + p\right) H \tilde{r}_A\, dt -  \frac{1}{2} A\left(\rho 
+ p\right) \dot{\tilde{r}}_A\, dt\,.
\end{equation}
In the regime where the apparent horizon radius remains nearly constant, such 
that $\dot{\tilde{r}}_A \ll 2H\tilde{r}_A$~\cite{Cai:2005ra,Sheykhi:2018dpn}, 
the second term becomes negligible. Under this approximation, the first law 
reduces to
\begin{equation}
\label{redfirstlaw}
T_h\, dS = A\left(\rho + p\right)\, H \tilde{r}_A\, dt\,.
\end{equation}

At this point, the standard Friedmann equations   can be readily derived by 
substituting the expression for the temperature $T_h$ and employing the 
Bekenstein-Hawking entropy-area relation. This procedure yields
\begin{eqnarray}
\dot{H} &=& -4\pi G\left(\rho + p\right)\,, \label{F1} \\[2mm]
H^2 &=& \frac{8\pi G}{3}\rho +\frac{\Lambda}{3}\,, \label{F2}
\end{eqnarray}
where $G = L_p^2$ denotes the  Newton gravitational constant, and $\Lambda$ is 
an integration constant that plays the role of the cosmological constant.

\subsection{Modified Friedmann equations through Barrow or Tsallis entropy}

We have shown that applying the first law of thermodynamics with 
Bekenstein–Hawking entropy to a homogeneous and isotropic Universe naturally 
reproduces the standard cosmological equations. This leads to the natural 
question of how cosmic dynamics would be modified if the underlying entropy 
deviates from the conventional holographic scaling. Specifically, by adopting 
the Barrow entropy formulation~\eqref{Barrow}, one obtains a modified set of 
Friedmann equations of the form (see appendix~\ref{AppA} for computational details)~\cite{Saridakis:2020lrg,Leon:2021wyx}
\begin{eqnarray}
\label{FM1}
\hspace{-2mm}\dot{H} &=&  -4\pi 
G\left(\rho_m+p_m+\rho_r+p_r+\rho_{\mathrm{DE}}+p_{\mathrm{DE}}\right),\\[2mm]
\hspace{-2mm}H^2 &=& \frac{8\pi G}{3}\left(\rho_m+\rho_r+\rho_{\mathrm{DE}}\right).
\label{FM2}
\end{eqnarray}
For later convenience, we have separated the contributions of \emph{matter} 
(baryons plus dark matter), \emph{radiation} and \emph{dark energy}. It is important to emphasize that the latter constitutes an effective component, as dark energy does not appear explicitly in Eqs.~\eqref{FM1} and~\eqref{FM2}. Instead, it emerges from geometric corrections to the horizon surface (i.e., $\Delta \neq 0$) induced by the Barrow entropy. For alternative Dark Universe models of similar geometric origin, one may refer to~\cite{Pal:2004ii} in the context of extended gravity theories with additional terms, and to~\cite{Luongo:2025iqq} regarding the potential contribution of gravitational metamaterials to the dark matter sector.

In the following analysis, we consider a matter component characterized by the 
equation of state $p_m = 0$, corresponding to pressureless dust. Under this 
assumption, the energy density and pressure associated with the effective dark  
energy component take the form given in Eqs.~\eqref{M1new}–\eqref{M2new}, which are rewritten below for convenience:
\begin{eqnarray}
\label{M1}
\rho_{\mathrm{DE}}&=& \frac{3}{8\pi G}\left\{
\frac{\Lambda}{3}+H^2\left[1-\frac{\beta\left(\Delta+2\right)}{2-\Delta}H^{
-\Delta}
\right]
\right\},\\[2mm]
p_{\mathrm{DE}} &=& -\frac{1}{8\pi G}\left\{\Lambda+2\dot 
H\left[1-\beta\left(1+\frac{\Delta}{2}\right)H^{-\Delta}\right] \right.
\nonumber\\
&&\left.
+3H^2\left[1-\frac{\beta\left(2+\Delta\right)}{2-\Delta}H^{-\Delta}\right]
\right\},
\label{M2}
\end{eqnarray}
where $\Lambda = 4CG\left(4\pi\right)^{\Delta/2}$, $C$ is  an integration 
constant and $\beta = \left(\pi/G\right)^{\Delta/2}$. 
As expected, in the case $\Delta=0$, we have $\rho_{\mathrm{DE}}=\Lambda/(8\pi G)$, 
$p_{\mathrm{DE}}=-\Lambda/(8\pi G)$, thus recovering the standard $\Lambda$CDM paradigm (see Eqs.~\eqref{parad1} and~\eqref{parad2}). As we mentioned in the Introduction, modified cosmology though Tsallis entropy can also be obtained from the same equations, under the identification $\Delta \rightarrow 2(\delta - 1)$.

In order to investigate the impact of the Barrow model on cosmic evolution, we now introduce the fractional energy densities $\Omega_i \equiv 8\pi G\rho_i/(3H^2)$ (the index $i = m, r$ corresponds  to matter and radiation, respectively), $\Omega_\Lambda \equiv \Lambda/(3H^2)$ and $\Omega_{\mathrm{DE}} \equiv {8\pi 
G\rho_{\mathrm{DE}}}/(3H^2)$. After some straightforward algebra, the evolution of the dimensionless Hubble parameter in terms of the redshift $z$ can be rewritten as (see Appendix for details)
\begin{eqnarray}
&&
\!\!\!\!\!\!\!\!\!\!\!\!\!\!\!\!\!\!\!\!
E(z) \equiv \frac{H(z)}{H_0}= 
\left[\frac{\bar{\beta}(2 - \Delta)}{2 + \Delta}\right]^{\frac{1}{2 - 
\Delta}} 
 \nonumber\\
&& \,
\cdot
\left[\Omega_{m0}\left(1 + z\right)^3 
+ \Omega_{r0}\left(1 + z\right)^4 + \Omega_{\Lambda 0}\right]^{\frac{1}{2 - 
\Delta}},
\label{NMFE}
\end{eqnarray}
where we have assumed the
standard scaling laws~\eqref{scalaw} for the matter and radiation components. Furthermore, we have defined $\bar{\beta}\equiv H_0^\Delta(G/\pi)^{\Delta/2}$ and denoted  the present-day value of each quantity by the subscript "0". The value of $\Omega_{\Lambda 0}$ can be fixed by requiring that the first Friedmann equation holds at the present time, i.e., by imposing the condition $E(0) = 1$ (see Eq.~\eqref{Omla0}).

At this point, the relation~\eqref{NMFE} can be simplified by noting that the prefactor $\frac{\bar{\beta}(2 - \Delta)}{2 + \Delta}$
is a multiplicative term that has no substantial impact on the dynamics of the cosmological evolution. 
Therefore, for the purposes of our 
subsequent analysis, we shall set it equal to unity. We thus arrive at
\begin{equation}
\label{Eznew}
    E(z) = \left[ E_{\Lambda}^{2}\left(z\right) \right]^{\frac{1}{2 - 
\Delta}}\,,
\end{equation}
where $E_\Lambda(z)$ is the dimensionless Hubble rate within the standard $\Lambda$CDM model (see the definition in Eq.~\eqref{Elambda}). 

\subsection{Understanding the effective dynamical dark energy}
\begin{figure*}[t]
\centering\includegraphics[width=0.7\textwidth]{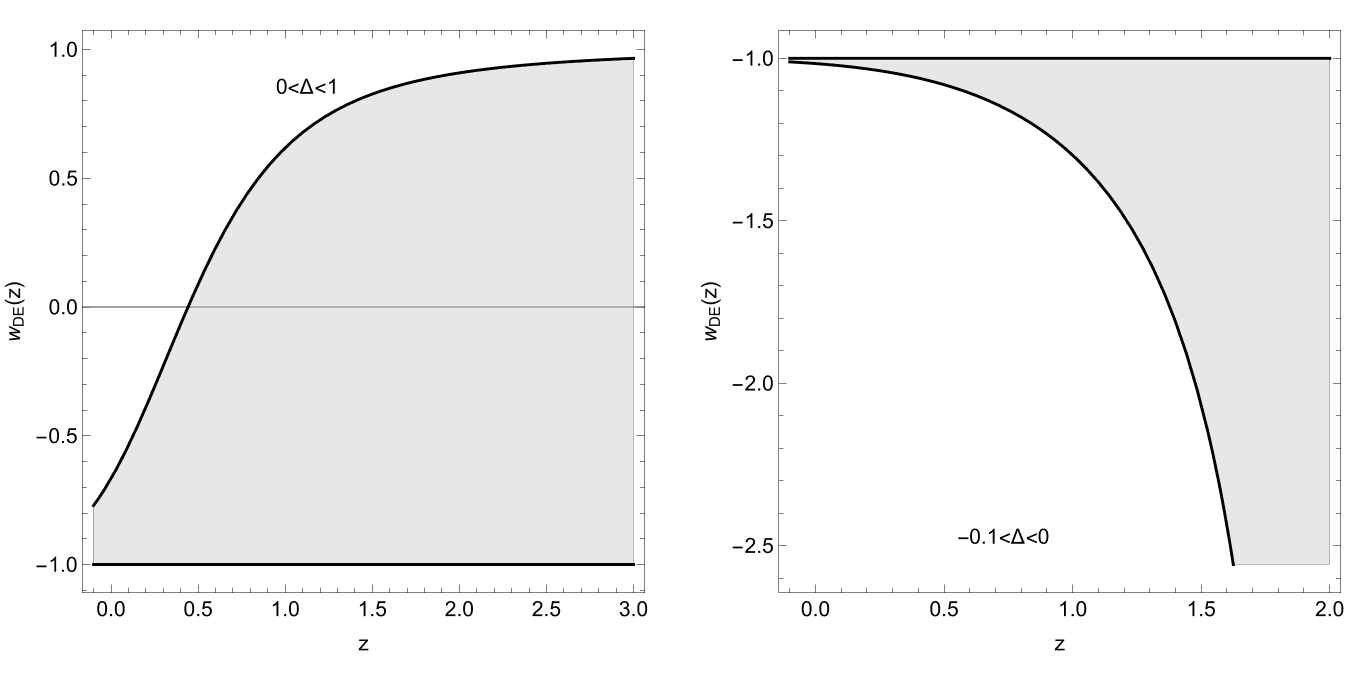}
\caption{{\it{Evolution of the effective dark energy equation-of-state 
parameter 
$w_{\mathrm{DE}}(z)$ for different values of the parameter $\Delta$. The left 
panel corresponds to $\Delta > 0$, while the right panel corresponds to $\Delta 
< 0$.    We have set $\Omega_{m0}=0.27$. The gray area represent the full set of curves, with boundaries the values of the $\Delta$ at the limits. }}}
\label{fig0}%
\end{figure*}

For the purposes of studying the late-time evolution of the Universe, we omit 
the radiation component in Eq.~\eqref{Eznew}, as its effects are essentially negligible. Under 
this assumption, we can write
\begin{equation}
\label{ez17}
E(z)\simeq\left[ \Omega _{m0}\left( 1+z\right)
^{3}+\left( 1-\Omega _{m0}\right) \right] ^{\frac{1}{2-\Delta }}\,,
\end{equation}
where we have used the condition~\eqref{Omla0} with $\frac{\bar{\beta}(2 - \Delta)}{2 + \Delta}=1$ and $\Omega_{r0}\simeq0$, as discussed above. 

In order to understand the equivalent dynamical dark energy model analogue 
related to  Barrow entropy, we parametrize the Hubble function as follows:
\begin{equation}
\label{newp}
E(z)\equiv \left[ \Omega _{m0}\left( 1\!+\!z\right) ^{3}+\left( 1\!-\!\Omega
_{m0}\right) \exp \left(\! 3\int \frac{1\!+\!w_{\mathrm{DE}}\left( z\right) 
}{1+z}dz\right)  
\right] ^{\frac{1}{2}}\!\!.
\end{equation}
Thus, by equating Eqs.~\eqref{ez17} and~\eqref{newp}, the effective equation-of-state parameter $w_{\mathrm{DE}}\left( z\right)$ is given by

\begin{widetext}
 \begin{equation}
 \label{eosp}
w_{\mathrm{DE}}\left( z\right) =\frac{\left( 1+z\right) 
^{-3}\Delta\hspace{0.5mm} \Omega
_{m0}+\left( 1+z\right) ^{-6}\left( \Delta -2\right) \left( 1-\Omega
_{m0}\right) }
{\left(2- \Delta \right) \left[\left(
1+z\right)^{-3}\left( 1-\Omega _{m0}\right)+\Omega _{m0}\right]
\left\{ \left( 1+z\right)
^{-3}-\Omega _{m0}\left\{ 1+\left[ \left( 1+z\right) ^{3}-1\right] \Omega
_{m0}\right\} ^{\frac{2}{\Delta -2}}\right\} }.
\end{equation}
\end{widetext}
It is straightforward to verify that $w_{\mathrm{DE}} \rightarrow -1$ as 
$\Delta \rightarrow 0$, thereby recovering the $\Lambda$CDM behavior in the 
limit where Barrow entropy reduces to the standard area-law scaling.

In Fig. \ref{fig0} we present the $w_{\mathrm{DE}}(z)$  range    for different 
values of the parameter $\Delta$. As we observe,  for $\Delta>0$, we have
$w_{\mathrm{DE}}(z)>-1$. On the other hand, phantom 
behaviour, namely $w_{\mathrm{DE}}(z)<-1$, occurs when $\Delta<0$.

\section{Observational Constraints}
\label{Obs} 
  
In this section we first describe the data that we use in order to constrain  
the modified Hubble function, and   we present the results on the scenario 
of modified cosmology through Barrow entropy. Additionally, we compare our 
results with  the Gaussian Process reconstruction of $w_{DE}$ for the 
DESI + Cosmic Microwave Background  (CMB) + Union3 combination 
\cite{DESI:2025fii}.

\subsection{Results}

In our analysis we use the following datasets.

\begin{itemize}
\item Observational Hubble Data (OHD): This data set includes 31 direct
measurements of the Hubble parameter from passive elliptic galaxies, known
as cosmic chronometers.\ The measurements for redshifts in the range $%
0.09\leq z\leq 1.965~$as summarized in \cite{cc1}.

\item Pantheon+ (SN/SN$_{0}$): This set includes 1701 light curves of 1550
spectroscopically confirmed supernova events within the range $%
10^{-3}<z<2.27~$\cite{Brout:2022vxf}. The data provides the distance modulus 
$\mu
^{obs}~$at~observed redshifts~$z$. We consider the Pantheon+ data with the
Supernova H0 for the Equation of State of Dark energy Cepheid host distances
calibration (SN$_{0}$) and without the Cepheid calibration (SN).

\item Baryonic acoustic oscillations (BAO): These data are provided by the
SDSS Galaxy Consensus, quasars and Lyman-$\alpha $ forests \cite{bbn0} and
the recent data from the DESI DR2 collaboration \cite{DESI:2025zgx}.
\end{itemize}

For the analysis, we employ COBAYA \cite{cob1,cob2}, with a 
custom  form for the Hubble function (\ref{ez17}), or the equivalent form (\ref{newp}), where the dynamical dark energy is expressed by Eq. (\ref{eosp}) and it depends on the parameters $\Delta$ and $\Omega_{m0}$. 
In other terms, we consider a $w\mathrm{CDM}$ cosmological model, where the dark energy density is not constant but evolves with a redshift-dependent equation of state $w_\mathrm{DE}(z)$. This function is not varied independently, but is implicitly determined by the parameters $\Delta$ and $\Omega_{m,0}$.
Moreover, we make use of the PolyChord nested sampler \cite{Handley:2015fda,poly2} which provides the Bayes
evidence. 

We consider the free parameters to be the energy density of the
dark matter  $\Omega _{DM}$, the Hubble constant $H_{0}$, the parameter $\Delta 
$, and the $r_{drag}$ which refers to the maximum distance sound waves could
travel in the early Universe before the drag epoch. For the energy density
of the baryons we consider the value provided by the Planck 2018
collaboration~\cite{planck}, $\Omega _{b0}=0.0486,$ while the radiation can
be omitted since its effects are neglected in the late universe.

We perform our analysis for different datasets. Specifically we consider the
datasets $\mathbf{D}_{1}:$SN+BAO, $\mathbf{D}_{2}:$SN+OHD+BAO, $\mathbf{D}%
_{3}:$SN$_{0}$+BAO and $\mathbf{D}_{4}:$SN$_{0}$+OHD+BAO. 

Barrow entropy is defined for $\Delta >0$. However, BAO data supports a
phantom behaviour in the past. Consequently, from Fig. \ref{fig0} we observe
that the effective equation of state parameter has a phantom behaviour in
the past for $\Delta <0$. Thus, for $\Delta $ we consider the prior $%
-0.5<\Delta <0.5$, this prior allow us to compare our results with the
previous analysis \cite{Anagnostopoulos:2020ctz}. 

As far as the priors  for the rest  of the free parameters are 
concerned, for the Hubble constant $H_0$ we consider $60<H_0<80$, for the 
energy density  $0.1<\Omega_m<0.5$ and for the 
$r_{drag}$ we consider $130<r_{drag}<160$.  

\begin{figure}[t]
\centering\includegraphics[width=0.48\textwidth]{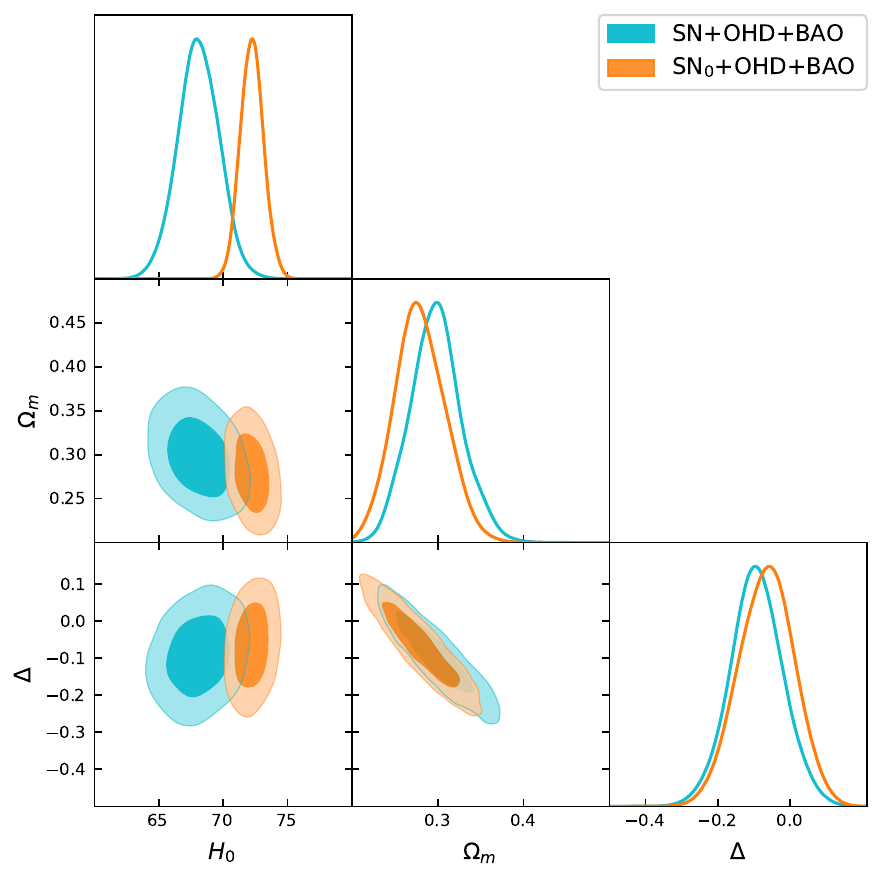}\caption{{
\it{Likelihood contours for the best-fit parameters of modified cosmology 
through Barrow entropy, for the datasets  
 $\mathbf{D}_{2}:$SN+OHD+BAO,   and $\mathbf{D}_{4}:$SN$_{0}$+OHD+BAO. The 
results for Tsallis exponent $\delta$ are obtained from those of $\Delta$ under 
the identification $\Delta \rightarrow 2(\delta - 1)$.   }}}%
\label{fig1}%
\end{figure}

The best fit cosmological parameters for the two different priors are
presented in Table \ref{data1}. Additionally, 
in Fig. \ref{fig1} we present the likelihood contours 
for the best-fit parameters of the scenario for the datasets 
$\mathbf{D}_{2}$ and $\mathbf{D}_{4}$.
As we see, the best fist parameter $\Delta$ 
is found to be negative, while the zero and the positive values are in the range 
of $2\sigma $ for the datasets $\mathbf{D}_{1},$ $\mathbf{D}_{2}$, $\mathbf{D}%
_{3}$ and $1\sigma $ for the dataset $\mathbf{D}_{4}\,$. This is different
from the results presented in \cite{Anagnostopoulos:2020ctz}, where the zero
value   was within the $1\sigma $. 

As emphasized in the Introduction, negative values of the anomalous dimension \(\Delta\) may carry meaningful physical implications beyond the original Barrow framework, suggesting a horizon surface with a fractal-like structure characterized by internal voids or porosity - such as sponge-like or porous geometries - rather than the sphereflake-type configurations typically associated with \(\Delta > 0\). From a theoretical standpoint, this behavior may also be intuitively linked to quantum gravitational effects. Indeed, in scenarios where quantum fluctuations dominate near the horizon, spacetime itself could acquire a nontrivial microscopic structure - akin to a porous or foam-like surface - leading to an effective reduction in dimensionality. Such features are reminiscent of models in quantum field theory where the renormalization group flow generates negative anomalous dimensions \cite{Dagotto:1989gp}.

Finally, note that for the $\mathbf{D}_{4}:$SN$_{0}$+OHD+BAO dataset, we obtain a value of $H_0 = 72.2_{-0.9}^{+0.9}$, which suggests a potential alleviation of the $H_0$ tension~\cite{DiValentino:2025sru}.However, it is important to emphasize that the present analysis does not include CMB observations. This should be kept in mind when interpreting the apparent alleviation of the Hubble tension, as our results are based exclusively on late-time and direct measurements, which typically prefer larger values of $H_0$. In contrast, including early-universe and indirect probes such as the CMB generally leads to lower estimates of $H_0$ (see, for instance,~\cite{DiValentino:2021izs}).

\begin{table}[tbp]
\centering
\caption{Cosmological constraints on modified cosmology through Barrow or 
Tsallis entropy. The results for Tsallis exponent $\delta$ are obtained from 
those of $\Delta$ under the identification $\Delta \rightarrow 2(\delta - 1)$.}
\resizebox{\columnwidth}{!}{
\begin{tabular}{ccccc}
\hline\hline
\textbf{BT Entropy} & $\mathbf{H}_0$ & $\mathbf{\Omega}_{m0}$ & 
$\mathbf{\Delta}$ & $\mathbf{\delta}$ \\
\hline
\textbf{SN+BAO} & $69.4_{-5.3}^{+3.9}$ & $0.297_{-0.033}^{+0.029}$ & 
$-0.096_{-0.075}^{+0.075}$ & $0.952^ {+0.038}_{-0.038}$ \\
\textbf{SN+OHD+BAO} & $68.1_{-1.6}^{+1.6}$ & $0.297_{-0.031}^{+0.028}$ & 
$-0.093_{-0.074}^{+0.074}$ & $0.954^{+0.037}_{-0.037}$ \\
\textbf{SN$_0$+BAO} & $73.6_{-0.99}^{+0.99}$ & $0.298_{-0.074}^{+0.074}$ & 
$-0.095_{-0.074}^{+0.074}$ & $0.953^{+0.037}_{-0.037}$ \\
\textbf{SN$_0$+OHD+BAO} & $72.2_{-0.9}^{+0.9}$ & $0.280_{-0.031}^{+0.028}$ & 
$-0.066_{-0.077}^{+0.077}$ & $0.967_{-0.039}^{+0.039}$ \\
\hline\hline
\end{tabular}
}
\label{data1}
\end{table}

Additionally, in order to compare the fitting efficiency and the behavior of 
the scenario with that of  $\Lambda $CDM paradigm, we fit the latter with the 
same datasets. In order to compare
the two different scenarios, we make use of the Akaike Information Criterion 
\cite{AIC} (AIC) and the Jeffrey's scale \cite{AIC2} for the Bayes evidence~$%
\log Z$. These criteria are used to compare models with different parametric
space. Indeed, modified cosmology through Barrow entropy has an additional free 
parameter. 

The AIC scale states that if $\left\vert \Delta \mathrm{AIC} 
\right\vert < 2$, the two models are statistically equivalent. When $2 \leq 
\left\vert \Delta \mathrm{AIC} \right\vert < 6$, there is weak evidence for a 
preference to the the model with the lower AIC. For $\left\vert \Delta 
\mathrm{AIC} \right\vert > 6$, the evidence becomes strong, while values 
exceeding $10$ indicate clear and decisive support for the model with the  
smaller AIC value. The quantity $\Delta \mathrm{AIC}$ is defined as the 
difference between the AIC values of two models, namely
\begin{equation}
\Delta \mathrm{AIC} = \mathrm{AIC}_1 - \mathrm{AIC}_2 = \chi^2_{min_1} - \chi^2_{min_2} + 2(N_1 - N_2),
\end{equation}
where $\chi^2_{min_i}$ is the minimum chi-squared value and $N_i$ is the number of free parameters of model $i$.

On the other hand, the difference in Bayesian evidence is defined as
\begin{equation}
\Delta \left( \ln Z \right) = \ln \frac{Z_2}{Z_1},
\end{equation}
where $\ln Z_i$ is the logarithm of the Bayesian evidence for model $i$. Then, Jeffrey's scale states that if
$\left\vert \Delta \left( \ln Z \right) \right\vert \ll 1$,  the models are 
indistinguishable. Furthermore, for $\left\vert \Delta \left( \ln Z \right) 
\right\vert < 1$, there exist a weak evidence in favor of the model with the 
higher $\ln Z$. A difference greater than $1$ indicates moderate evidence, 
while $\left\vert \Delta \left( \ln Z \right) \right\vert > 2.5$ suggests 
strong evidence. Finally, for $\left\vert \Delta \left( \ln Z \right) 
\right\vert > 5$, the evidence is very strong.

In Table \ref{data2} we present the difference of the statistical parameters
between the two models. It follows that the entropic model leads to a $\chi
_{\min }^{2}$ which is slightly smaller for the datasets that we
investigate. However, from the AIC and Jeffrey's scale for the {$\log \left( 
\frac{\mathbf{Z}%
_{\Lambda }}{\mathbf{Z}}\right) $ we infer that there is a Weak/Moderate
evidence in favor of the $\Lambda$CDM.

\begin{table}[!] \centering%
\caption{Comparison of Barrow-Tsallis Entropy with the $\Lambda$CDM paradigm.}%
\begin{tabular}{cccc}
\hline\hline
\textbf{BT Entropy} & $\mathbf{\chi }_{\min }^{2}-\mathbf{\chi }%
_{\Lambda \min }^{2}$ & $\mathbf{AIC}-\mathbf{AIC}_{\Lambda }$ & {$\log
\left( \frac{\mathbf{Z}_{\Lambda }}{\mathbf{Z}}\right) {\,}$} \\ \hline
\textbf{SN+BAO} & $-0.4$ & $+1.6$ & $+0.65$ \\ 
\textbf{SN+OHD+BAO} & $-0.5$ & $+1.5$ & $+1.04$ \\ 
\textbf{SN}$_{0}$\textbf{+BAO} & $-0.6$ & $+1.4$ & $+0.85$ \\ 
\textbf{SN}$_{0}$\textbf{+OHD+BAO} & $+0.5$ & $+2.5$ & $+1.57$ \\ 
\hline\hline
\end{tabular}%
\label{data2}%
\end{table}%

\subsection{Comparison with Gaussian Process reconstruction of $w_{DE}$ for the 
DESI + CMB + Union3 combination}

In order to obtain a more transparent picture for the results 
obtained in Table~\ref{data1}, and in particular to investigate the 
implications of a negative Barrow exponent on the cosmic evolution, we now 
compare the predictions of our model for the effective dark energy 
equation-of-state parameter with the reconstruction proposed 
in~\cite{DESI:2025fii}, based on DESI + CMB + Union3 measurements. The latter 
was performed using Gaussian Process (GP) regression, a non-parametric 
statistical tool widely employed in various fields to reconstruct smooth 
functions from noisy data without assuming a specific functional form (see, for 
example~\cite{Calderon:2022cfj}).

The results are illustrated in Fig.~\ref{lastfig}, where the blue shaded bands 
represent the confidence intervals at various significance levels for the 
GP-reconstructed $w_{DE}$.
Several points deserve to be emphasized: first, as shown in Sec.~\ref{ModCosm}, the present parametrization of the Barrow-Tsallis model predicts that there is no phantom crossing of the equation-of-state parameter, with $\omega_{DE} > -1$ for $\Delta > 0$ (left panel in Fig.~\ref{fig0}) and $\omega_{DE} < -1$ for $\Delta < 0$ (right panel in Fig.~\ref{fig0}). On the other hand, the GP reconstruction based on DESI + CMB + Union3 measurements appears to favor models that exhibit a phantom crossing, although alternatives lacking this feature cannot be completely ruled out (see~\cite{DESI:2025fii} for more details). Furthermore, for $\Delta = -0.093^{+0.074}_{-0.074}$, obtained through SN+OHD+BAO data, Fig.~\ref{lastfig} illustrates that the set of curves predicted by our model (red bands) significantly overlaps with the GP reconstruction at the $1\sigma$ level.

It is important to note, however, that our analysis, unlike the one in~\cite{DESI:2025fii}, is based exclusively on late-time observational probes and does not incorporate early-universe data such as the CMB. To enable a more comprehensive comparison with the reconstruction in~\cite{DESI:2025fii}, an extension of our model to early cosmological times would be required. In this context, additional physical effects not included in the present treatment (such as the radiation component, scale-dependent modifications of gravity and other high-redshift corrections) could become significant and affect the evolution of $\omega_{DE}$. These aspects deserve further investigation and will be addressed in a forthcoming analysis.

\begin{figure}[t]
\centering\includegraphics[width=0.45\textwidth]{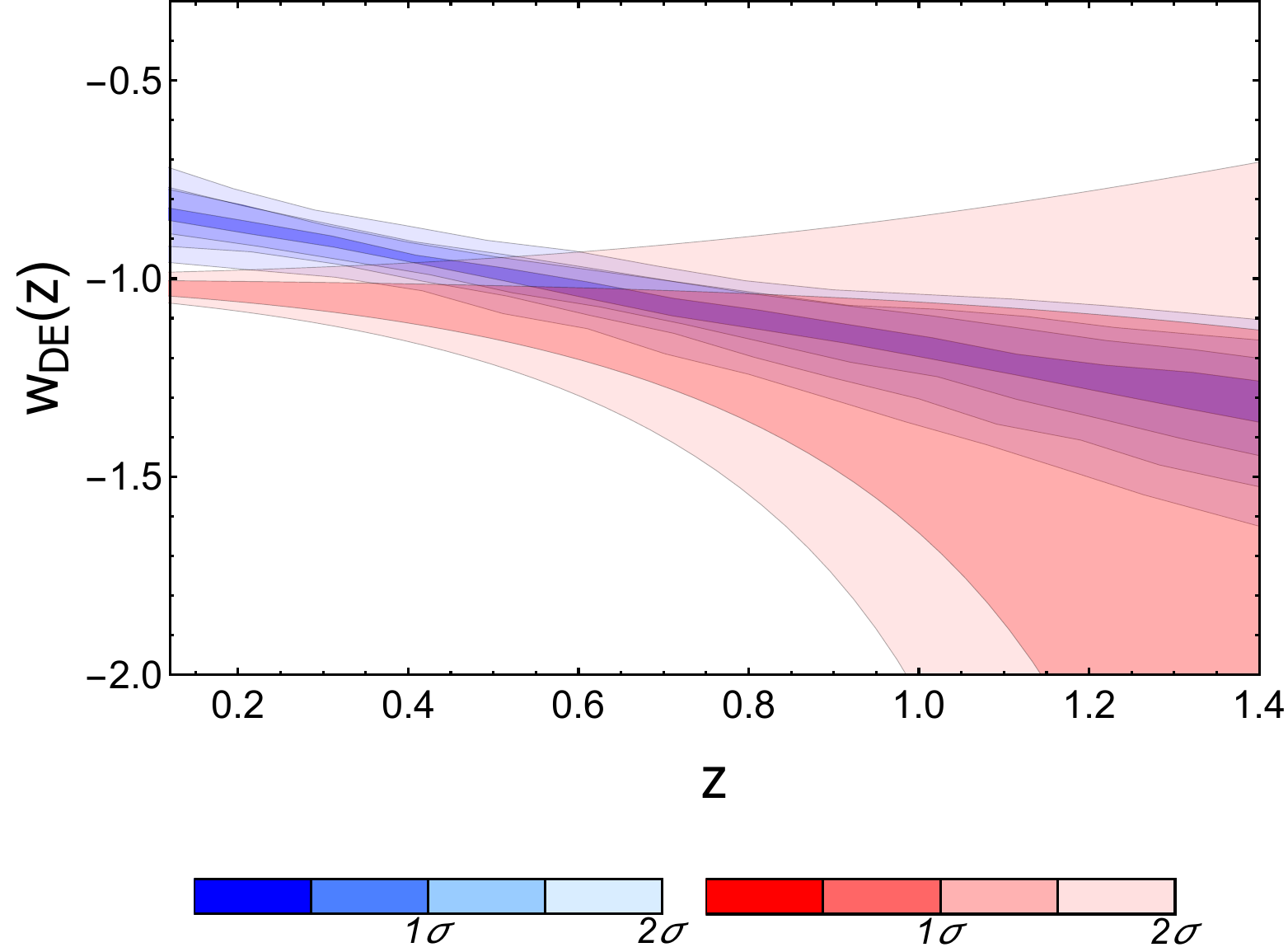}\caption{{
\it{Gaussian Process reconstruction of $w_{DE}$ for the DESI + CMB + Union3 
combination (blue bands)~\cite{DESI:2025fii} versus BT $w_{DE}$ for the SN+OHD+BAO combination (red bands). The shaded regions illustrate confidence intervals at various 
significance levels.
}}}%
\label{lastfig}%
\end{figure}

\section{Conclusions and Outlook}
\label{Conc}

A well-established conjecture in theoretical physics suggests a deep connection 
between gravity and thermodynamics. Within cosmological contexts, this 
correspondence implies that the standard Friedmann equations governing the 
Universe expansion may be derived from the first law of thermodynamics using 
the standard Bekenstein-Hawking entropy. On the other hand, Barrow entropy is 
an one-parameter modification of Bekenstein-Hawking entropy, arising from   
quantum-gravitational effects that induce a  fractal structure on the surface 
of black holes. Similarly, Tasllis entropy is an one-parameter modification arising within the framework of nonextensive statistical mechanics. 
Hence, if one applies the gravity-thermodynamics conjecture, 
but using the modified Barrow or Tsallis entropy then one obtains modified Friedmann 
equations and thus a modified cosmological scenario. 

In this work we use observational data from Supernova Type Ia (SNIa), Cosmic 
Chronometers (CC), and Baryonic acoustic oscillations (BAO), including the 
recently released DESI  DR2 data, in order to extract constraints on the 
scenario of modified cosmology through Barrow and Tsallis entropy. Firstly, we examined the 
behavior of the effective dark-energy equation-of-state parameter according to 
the values of the Barrow exponent $\Delta$, or the Tsallis exponent $\delta=1+\Delta/2$, showing that it lies in the 
quintessence regime for positive $\Delta$, while it lies in the phantom regime 
for negative $\Delta$. Then we performed the full observational confrontation,
focusing on the constraints on $\Delta$,  on the current matter density 
parameter $\mathbf{\Omega }_{m0}$ and on the current Hubble function value 
$H_0$.

As we showed,   the best-fit  value for $\Delta$ 
is found to be negative, while the zero value, which corresponds to 
$\Lambda$CDM paradigm,   is allowed only   in the range of $2\sigma $ for the 
datasets $\mathbf{D}_{1}:$SN+BAO, $\mathbf{D}_{2}:$SN+OHD+BAO, $\mathbf{D}%
_{3}:$SN$_{0}$+BAO  and in the range of $1\sigma $ for the dataset 
$\mathbf{D}_{4}\,$. Additionally, for the case of the   
$\mathbf{D}_{4}:$SN$_{0}$+OHD+BAO dataset, we obtained a value of $H_0= 
72.2_{-0.9}^{+0.9}$, which may potentially alleviate the $H_0$ 
tension. This conclusion is further supported by the fact that, as discussed in 
\cite{Heisenberg:2022gqk}, a mechanism for  alleviating the $H_0$ tension is 
that the effective dark energy component exhibits a phantom-like behavior at 
some redshift $z$. This feature is indeed realized in our model for $\Delta < 
0$. Clearly, a more definitive assessment requires a dedicated observational 
analysis, including a comparison with data
from multiple cosmological surveys, 
such as Planck \cite{Planck:2018vyg}, SH0ES \cite{Riess:2021jrx} and H0LiCOW \cite{H0LiCOW:2019pvv}, which provide complementary measurements of the Hubble constant across different redshift regimes.
A study along this direction is planned as a future extension of the present work. Some results can be found in \cite{Basilakos:2023kvk}.
 Finally,   we 
compared our 
results with  the Gaussian Process reconstruction of the dark-energy 
equation-of-state parameter $w_{DE}$ for the DESI +  CMB  + Union3 datasets, 
showing that the corresponding likelihood contours lie within the   
boundaries of the $w_{DE}$-region  predicted by our model.

By applying information criteria such as the  Akaike Information 
Criterion and the   Bayes evidence, we then compared the fitting efficiency of the scenario at hand with $\Lambda$CDM  cosmology, showing that the latter is 
slightly favoured. Finally, we investigated the compatibility of our model with the Gaussian Process reconstruction of the effective dark energy equation-of-state parameter based on DESI + CMB + Union3 data. We found that, in the case of negative $\Delta$, the predictions of the Barrow-Tsallis model show overlap with this  reconstruction at the $1\sigma$ level, particularly at low redshift.

Recent observational datasets from  DESI  DR2 collaboration seem to favour 
dynamical dark energy. Hence, it is crucial to investigate scenarios that 
deviate from $\Lambda$CDM paradigm in a dynamical way. Modified cosmologies 
through Barrow, Tsallis and other extended entropies lie in this category of dynamical 
dark energy. Thus, they may offer an alternative   for the description of 
nature.

\appendix
\section{Modified Cosmology from Barrow/Tsallis Entropy}
\label{AppA}

In this appendix, we present computational details of the derivation of the generalized Friedmann equations in the context of the Barrow/Tsallis entropy model. The procedure, based on the gravity–thermodynamics conjecture and the assumptions described in Sec.~\ref{ModCosm}, closely follows the standard derivation. The only difference lies in the form of the horizon entropy, which in the present case is given by Eq.~\eqref{Barrow}, instead of the usual Bekenstein–Hawking scaling. Accordingly, the differential of the entropy takes the form \begin{equation}
dS_B=\left(\frac{4\pi}{A_0}\right)^{1+\frac{\Delta}{2}}\left(2+\Delta\right)\tilde r_A^{1+\Delta}\hspace{0.3mm}\dot{\tilde r}_A dt,
\end{equation}
where we have used the relation $A=4\pi\tilde r_A^2$. By substituting this expression into the first law of thermodynamics~\eqref{redfirstlaw} applied to the cosmological horizon, and using the Hawking definition of the horizon temperature, $T_h = 1/(2\pi\tilde{r}_A)$, we arrive at
\begin{equation}
\label{dotHapp}
\dot H=-\frac{2^{1-\Delta}\pi^{1-\frac{\Delta}{2}}A_0^{1+\frac{\Delta}{2}}}{2+\Delta}\hspace{0.2mm}{H^\Delta}\left(\rho+p\right),
\end{equation}
where we have expressed the horizon radius in terms of the Hubble parameter 
through $\tilde{r}_A = 1/H$, which holds for a spatially flat, homogeneous and 
isotropic FRW universe. The above relation provides the implicit form of the 
modified Friedmann equation for the derivative of the Hubble rate.

Finally, upon employing the continuity equation~\eqref{cont} and carrying out the integration, we obtain the other modified Friedmann equation
\begin{equation}
\left(\frac{2+\Delta}{2-\Delta}\right)H^{2-\Delta}=\left[\frac{\left(4\pi\right)^{1-\frac{\Delta}{2}}}{6}\hspace{0.2mm}\rho+\frac{C}{3}\right]A_0^{1+\frac{\Delta}{2}}\,,
\label{Happ}
\end{equation}
where $C$ is the integration constant with dimensions $[L^{-4}]$. It is worth noting that the above relation remains valid as long as \( \Delta \neq 2 \), which is naturally fulfilled in the original Barrow scenario (where \( \Delta \leq 1 \)), as well as in the generalized framework considered here, which extends the Barrow model by also allowing negative values of \( \Delta \).

To gain more physical insight into the meaning of the corrections introduced by Barrow entropy, it is convenient to rewrite Eqs.~\eqref{dotHapp} and~\eqref{Happ} in the equivalent form
\begin{eqnarray}
\dot{H} &=&  -4\pi 
G\left(\rho+p+\rho_{\mathrm{DE}}+p_{\mathrm{DE}}\right),\\[2mm]
\hspace{-2mm}H^2 &=& \frac{8\pi G}{3}\left(\rho+\rho_{\mathrm{DE}}\right),
\label{Happendix}
\end{eqnarray}
where $\rho$ and $p$ denote the energy density and pressure of the perfect fluid filling the Universe (which generally includes both matter and radiation) while $\rho_{\mathrm{DE}}$ and $p_{\mathrm{DE}}$ refer to an effective dark energy sector that emerges from the modified holographic horizon description in the Barrow entropy framework. They are defined as
\begin{eqnarray}
\label{M1new}
\rho_{\mathrm{DE}}&=& \frac{3}{8\pi G}\left\{
\frac{\Lambda}{3}+H^2\left[1-\frac{\beta\left(\Delta+2\right)}{2-\Delta}H^{
-\Delta}
\right]
\right\},\\[2mm]
p_{\mathrm{DE}} &=& -\frac{1}{8\pi G}\left\{\Lambda+2\dot 
H\left[1-\beta\left(1+\frac{\Delta}{2}\right)H^{-\Delta}\right] \right.
\nonumber\\
&&\left.
+3H^2\left[1-\frac{\beta\left(2+\Delta\right)}{2-\Delta}H^{-\Delta}\right]
\right\},
\label{M2new}
\end{eqnarray}
respectively, where we have used $A_0=4G$ and defined 
\begin{equation}
\Lambda\equiv4CG\left(4\pi\right)^{\Delta/2}, \qquad  \beta \equiv \left(\pi/G\right)^{\Delta/2}\,.
\end{equation}
It is straightforward to verify that, for $\Delta = 0$, one has $\rho_{\mathrm{DE}} = \Lambda / (8\pi G)$ and $p_{\mathrm{DE}} = -\Lambda / (8\pi G)$. Therefore, the modified Friedmann equations~\eqref{M1new}-\eqref{M2new} reduce to those of the standard $\Lambda$CDM paradigm, namely
\begin{eqnarray}
\label{parad1}
  \dot H\big|_{\Delta=0}&=&-4\pi G\left(\rho+p\right),\\[2mm]
    H^2\big|_{\Delta=0}&=&\frac{8\pi G}{3}\rho+\frac{\Lambda}{3}\,,
    \label{parad2}
\end{eqnarray} 
which reveal the physical interpretation of the integration constant $\Lambda$ 
as the cosmological constant.

For our purposes of studying the effects of the Barrow model on cosmic evolution, let us now introduce the fractional energy density parameters
\begin{eqnarray}
&&\label{Omegr}
\!\!\!\!
\Omega_i \equiv \frac{8\pi G\rho_i}{3H^2}\,, 
\nonumber\\
&&\!\!\!\!\Omega_\Lambda \equiv  \frac{\Lambda}{3H^2}\,,
\nonumber\\
&&\!\!\!\!\Omega_{\mathrm{DE}} \  \equiv  \ \frac{8\pi 
G\rho_{\mathrm{DE}}}{3H^2} \ = \ 1 + \Omega_\Lambda - 
\frac{\beta\left(\Delta+2\right)H^{-\Delta}}{2-\Delta}\,,\ \ \ \ \
\end{eqnarray}
where the index $i = m, r$ corresponds  to matter and radiation, respectively. 
From Eq.~\eqref{Happendix}, it follows that the dimensionless form of the 
cosmological equation becomes
\begin{equation}
\Omega_m + \Omega_r + \Omega_\Lambda = 
\frac{\beta\left(\Delta+2\right)H^{-\Delta}}{2-\Delta}\,,
\label{Dimensionless}
\end{equation}
which can be written in the equivalent form
\begin{equation}
\Omega_m + \Omega_r + \Omega_{\mathrm{DE}} = 1\,.
\end{equation}
Here, $\Omega_m$ accounts for the total matter content of the Universe, 
including both dark matter and ordinary (baryonic) matter, i.e. $\Omega_m = 
\Omega_b + \Omega_{\mathrm{DM}}$.

Assuming that the matter and radiation components evolve according to their 
standard scaling laws, we have
\begin{equation}
\label{scalaw}
\rho_m = \rho_{m0}(1 + z)^3\,, \quad\,\, \rho_r = \rho_{r0}(1 + z)^4\,,
\end{equation}
where the redshift is defined as $z(t) = [1 - a(t)] / a(t)$, and the subscript 
"0" denotes the present-day value of the corresponding quantity. 
Hence, we acquire 
\begin{eqnarray}
&&
\!\!\!\!\!\!\!\!\!\!\!\!\!\!\!\!\!\!\!\!
E(z) \equiv \frac{H(z)}{H_0}= 
\left[\frac{\bar{\beta}(2 - \Delta)}{2 + \Delta}\right]^{\frac{1}{2 - 
\Delta}} 
 \nonumber\\
&& \,
\cdot
\left[\Omega_{m0}\left(1 + z\right)^3 
+ \Omega_{r0}\left(1 + z\right)^4 + \Omega_{\Lambda 0}\right]^{\frac{1}{2 - 
\Delta}},
\label{NMFEapp}
\end{eqnarray}
where we have defined $\bar{\beta}=H_0^\Delta(G/\pi)^{\Delta/2}$. Note that the 
value of $\Omega_{\Lambda 0}$ can be determined by imposing the first Friedmann 
equation at present time, $E(0) = 1$. Substituting into Eq.~(\ref{NMFEapp}) 
this yields
\begin{equation}
\label{Omla0}
\Omega_{\Lambda 0} = \left(\frac{2 + \Delta}{2 - 
\Delta}\right)\frac{1}{\bar{\beta}} - \Omega_{m0} - \Omega_{r0}\,.
\end{equation}
We finally observe that, for $\Delta = 0$, Eqs.~\eqref{NMFEapp} and~\eqref{Omla0} consistently reduce to
\begin{eqnarray}
\label{ezapp}
\nonumber
E(z)\big|_{\Delta=0}\hspace{-1mm}&=&\hspace{-1mm}E_\Lambda(z)\equiv \sqrt{\Omega_{m0}\left(1 + z\right)^3 + \Omega_{r0}\left(1 + 
z\right)^4 + \Omega_{\Lambda 0}}\,, \\
\label{Elambda}\\[2mm]
\Omega_{\Lambda 0}\big|_{\Delta=0} &=& 1 - \Omega_{m0} - \Omega_{r0}\,,
   \label{om0app}
\end{eqnarray}
respectively, thereby recovering the standard Hubble rate evolution and flatness 
condition of the $\Lambda$CDM cosmology.

\acknowledgments 
The authors are grateful to the anonymous Referee for the valuable comments on the original manuscript.  The research of GGL is supported by the postdoctoral fellowship program of the 
University of Lleida. GGL and ENS gratefully acknowledge  the contribution of 
the LISA 
Cosmology Working Group (CosWG), as well as support from the COST Actions 
CA21136 - \textit{Addressing observational tensions in cosmology with 
systematics and fundamental physics (CosmoVerse)} - CA23130, \textit{Bridging 
high and low energies in search of quantum gravity (BridgeQG)} and CA21106 -  
\textit{COSMIC WISPers in the Dark Universe: Theory, astrophysics and 
experiments (CosmicWISPers)}. AP thanks the support of VRIDT through 
Resoluci\'{o}n VRIDT No. 096/2022 and Resoluci\'{o}n VRIDT No. 098/2022. AP was 
Financially supported by
FONDECYT 1240514 ETAPA 2025.

\bibliography{Bib}

\end{document}